\documentclass[11pt]{article}
\usepackage{graphicx}
\usepackage{hyperref}
\usepackage{verbatim}
\usepackage{amsmath, amsthm, amssymb, multicol, color}
\usepackage{float}
\usepackage{cite}
\usepackage{lscape}
\usepackage{listing}
\usepackage{subcaption}
\usepackage{mwe}
\usepackage{tikz}
\usetikzlibrary{matrix}
\usepackage{titlesec}
\usepackage[hmargin=1in,vmargin=1in]{geometry}
\usepackage{setspace}
\title{Renormalization Group as a Koopman Operator}
\begin{document}

\section*{\centering Renormalization Group as a Koopman Operator}

\centering
William T. Redman

\centering
University of California, Santa Barbara

\centering
wredman@ucsb.edu

\abstract{Koopman operator theory is shown to be directly related to the renormalization group. This observation allows us, with no assumption of translational invariance, to compute the critical exponents $\eta$ and $\delta$, as well as ratios of critical exponents, of classical spin systems from single observables alone. This broadens the types of problems that the renormalization group framework can be applied to and establish universality classes of. In addition, this connection may allow for a new, data-driven way in which to find the renormalization group fixed point(s), and their relevant and irrelevant directions.  }

\section{\label{sec:level1}Introduction \protect\\ }

Understanding the behavior of multi-component complex systems is a central problem in all areas of science. In physics, the study of phase transitions and critical phenomena led to the development of the renormalization group (RG), one of the most successful tools in theoretical physics \cite{Kadanoff1966, Wilson1971A, Wilson1971B, Wilson1975}. By turning disparate physical problems into dynamical systems problems, it enabled a coherent understanding of the universality of critical exponents and allowed a controlled, albeit difficult, manner in which to compute them. 

Given its success, there have been attempts in applying the RG framework to problems beyond traditional physics. A recent example was the discovery of an exact mapping between the RG and deep neural networks (DNNs) based on Restricted Boltzmann machines (RBMs) \cite{Mehta2014}. However, a challenge in directly importing ideas and techniques from RG theory to these systems is that they do not have the same symmetries as the problems that RG theory was developed on have (e.g. translational invariance). Therefore, the extension of the RG framework to compute the critical exponents, and thereby define the universality classes, of non-symmetric systems is an important open question that likely requires a new approach. 


A recent advance in the study of nonlinear dynamical systems is Koopman operator theory (KOT). In addition to being based on a foundation of rigorous mathematical theory, there are a number of effective, data-driven algorithms inspired by KOT that extract the dynamics underlying a wide range of physical systems. KOT has been indirectly linked with the RG in several different fields. First, both the RG and KOT have been used to successfully discover the dynamics of partial differential equations \cite{Chen1994, Chen1996, Kutz2016}. Second, the RG has been used to extend principal component analysis (PCA) and perform dimensionality reduction \cite{Bradde2017}, while KOT has been shown to be related to PCA \cite{Brunton2016JNeuro, Klus2018}. Third, recent work applying normal form theory to the RG equations gave a unified approach to characterize the nonlinear generalizations of scaling functions \cite{Raju2019}. KOT has connections to normal form theory \cite{Mauroy2016, Mohr2016, Mezic2019}. Finally, the RG has been shown to have an exact mapping to RBM DNNs \cite{Mehta2014}, and KOT has been used to develop novel machine learning techniques \cite{Kawahara2016, Takeishi2017a, Takeishi2017b}. 

These connections between the RG and KOT motivated us to look more closely at the two. Does there exist a direct link between them? As we show below, \textit{the RG is, in fact, a Koopman operator}.

This paper is organized as follows. We begin by providing a brief overview of basic KOT. This will allow us to show that, by definition, the RG is a Koopman operator. We include references to a number of recent theoretical and applied research, which provide more detail about KOT, for the interested reader. We then use a KOT inspired algorithm to compute two critical exponents, and two ratios of critical exponents, of the Ising and three-state Potts models in 2D. We show that our algorithm requires only measuring a single observable, and does not rely on explicit calculations of how applying the RG changes coupling constants. We compare these results to the standard Monte Carlo RG (MCRG) approach. We find that our method performs similarly, while requiring less information about the exact RG flow, and is considerably faster. We end by outlining possible uses of this Koopman RG. In particular, we suggest that it can be used to compute the universality classes of more complex systems, such as RBM DNNs and spin glasses. Finally, we note that, by using more advanced KOT methods, we may be able to compute the RG fixed point(s), and their relevant (unstable) and irrelevant (stable) directions, in a data-driven way. 

\pagebreak

\section{\label{sec:level1} KOT and its connection to the RG operator\protect\\ }

KOT is a spectral dynamical systems theory that was first developed by Bernard Koopman in 1931 in the context of classical mechanics \cite{Koopman1931}, and then later expanded upon by Koopman and John von Neumann in 1932 \cite{Koopman1932}. It has seen a great increase in attention over the past two decades as a wave of new data-driven methods \cite{Mezic2013, Williams2014, Fujii2019, Takeishi2019, Mezic2019} and underlying mathematical theory \cite{Mezic2019, Mezic2019b, Mezic2005} has allowed it to be applied to the dynamics of fluids \cite{Rowley2009, Budisic2012, Brunton2016, Arbabi2017}, and to the study of power grids \cite{Susuki2016, Korda2018}, logistics \cite{Hogg2019}, urban insurgency \cite{Fonoberova2018}, and building energy \cite{Georgescu2015}. 

The key insight in KOT is that there exists an infinite dimensional linear operator, the Koopman operator (which is related to the composition operator), whose spectrum provides information on the dynamics of nonlinear systems. The Koopman operator, $U^t$, is defined, to be the time evolution operator of a given observable $\mathbf{g}$
\begin{equation}
\label{KOT eq}
    U^t \mathbf{g}\left( \mathbf{x}_0\right) = \mathbf{g}\left( \mathbf{S}^t \left(\mathbf{x}_0\right) \right)
\end{equation}
where $\mathbf{S}^t$ is the dynamics that act on $\mathbf{g}$ \cite{Mezic2019}. Time here can either be discrete or continuous. 

The block spin RG is defined to be a map in the infinite dimensional space of possible Hamiltonians with coupling constants $\left(K_1, K_2, ... \right)$, which we refer to as K-space from now on \cite{Goldenfeld2005}. While K-space is infinite dimensional, because we are ultimately interested in real systems that are assumed to have a finite interaction length, we will restrict ourselves to assuming that K-space is finite dimensional, with $d_K$ dimensions, and with a ``vanilla'' $L^2$ function space on it. This additionally allows us to avoid problems in nonlinear functional analysis that come with considering an infinite dimensional space. 

The block spin RG transformation, $R_b$, acting on a Hamiltonian with coupling constants $\mathbf{K}_0 = (K_1, K_2, ..., K_{d_K})$, is equivalent to a coarse grained Hamiltonian with new coupling constants. This can be done $n$ times

\begin{equation}
\label{RG KOT eq}
    R_b^n H(\mathbf{K}_0) = H(T^n(\mathbf{K}_0))
\end{equation}
where $T$ is the transformation from one point in K-space to another, following the chosen blocking procedure. Note that, when considering the evolution of $H(\mathbf{K}_0)$ in this way (with the RG iteration number acting as a ``time''), we are defining a dynamical system. It is from this that can connect the RG to KOT. 

Eqs. \ref{KOT eq} and \ref{RG KOT eq} immediately imply that, by definition, the RG is a Koopman operator in K-space. Because the Hamiltonian defines the value of all of the observables of interest (such as magnetization), we can apply KOT methods to their measured quantities to gain information about $R_b$. Although we only consider the block spin RG here, the Wilsonian (momentum space) RG, which integrates over continuous degrees of freedom, is a continuous Koopman operator.

The finite section method (also commonly known as the Galerkin projection) is a simple algorithm that is frequently used to compute the Koopman operator from time series data \cite{Mezic2019, Williams2014}. It finds the approximate Koopman operator, $\tilde{U}$, from the data matrix $F$, which is comprised of the first $n$ time points of $m$ observables $\mathbf{g}_1, ..., \mathbf{g}_m$. That is $F = \Big[\mathbf{g}_1 | ... | \mathbf{g}_m \Big]$. In particular, $\tilde{U}$ is given by 
\begin{equation}
\label{FSM eq}
    \tilde{U} = F^+ F'
\end{equation}
where $F'$ is the data matrix shifted one time step forward and $F^+$ is the Moore-Penrose pseudoinverse of $F$ (i.e. $F^+ = (F^\dagger F)^{-1} F^\dagger$). In practice, a single observable is often used to populate the data matrix, with each column being time delayed from the others \cite{Takens1981}. This allows for a rich, informative spectrum even from a single observable.

\section{\label{sec:level1}Results \protect\\ }

Having found that the RG is a Koopman operator in K-space, we explored whether we could successfully apply tools from KOT to calculate the critical exponents of the 2D Ising model and the 2D three-state Potts model. 

These classical spin models have a Hamiltonian of the form
\begin{equation}
\label{Spin Hamiltonian}
    H(\mathbf{K}) = - \sum_i K_1 s_i - \sum_{\langle i, j \rangle} K_2 s_i s_j - ...
\end{equation}
where the $\{s_i\}$ are the spins of the system, and the $K_i$ are the strengths of the different interaction types (e.g. $K_2$ is the strength of the nearest neighbor coupling). In these models, $K_1$ is interpreted as an external magnetic field, and is often referred to as $h$ instead. 

We considered the magnetization, $m$, which scales as 
\begin{equation}
\label{beta eq}
m \sim t^{\beta}
\end{equation}
near the critical manifold, where $t$ is the difference between the critical temperature, $T_c$, and actual temperature (i.e. $t = (T_c - T)/T_c$) \cite{Goldenfeld2005}. Because the magnetization is a function of the Hamiltonian, there exists a Koopman operator, $U_m$, that evolves the magnetization in RG iteration time. This is a proxy for the block spin RG, $R_b$, and its spectrum is related to $\beta$ (see below). 

To numerically approximate $U_m$, we first equilibrated spin systems using standard Monte Carlo approaches \cite{Newman1999a}. We then performed block spin renormalizations, with $b^2$ spins in each block, $n_{\text{R}}$ times, measuring the magnetization after each blocking (see Supplemental Material for more details about our procedure). This gave $n_{\text{R}} + 1$ values of $m$. We used the finite section method, Eq. \ref{FSM eq}, to get
\begin{equation}
\label{Mag KO eq}
    \tilde{U}_m = \left(m_0, ..., m_{n_{\text{R}} - 1}\right)^{+} \left(m_1, ..., m_{n_{\text{R}}}\right)^\dagger
\end{equation}
where $m_j$ is the magnetization of the spin system after $j$ block spin renormalizations. 

To see how the spectrum of $\tilde{U}_m$ is related to $\beta$, note that near the critical temperature 
\begin{equation}
\label{beta nu eq}
m_k \sim t_k^{\beta} \sim b^{\beta k/ \nu} t^{\beta} 
\end{equation}
where $\nu$ is the critical exponent related to the change in $t$ from applying the block spin RG (i.e. $t_{k} \sim b^{k / \nu} t $) and to the correlation length (i.e. $\xi \sim t^{-\nu}$) \cite{Goldenfeld2005}. Therefore,
\begin{equation}
\label{b beta nu eq}
    \begin{split}
    \left(m_1, ..., m_{n_{\text{R}}}\right) & \sim \left(b^{\beta / \nu}t^{\beta}, ..., b^{\beta n_{R} / \nu} t^{\beta}  \right)\\
    & = b^{\beta / \nu} \left(t^{\beta}, ..., b^{\beta (n_{R} - 1) / \nu} t^{\beta}  \right)\\
    & = b^{\beta / \nu} \left(m_0, ..., m_{n_{\text{R}} - 1}\right)
    \end{split}
\end{equation}
and so, by Eq. \ref{Mag KO eq}, we have that 
\begin{equation}
\label{Mag KO beta nu eq}
\tilde{U}_m \sim b^{\beta / \nu}
\end{equation}
Taking $\log_b$ of Eq. \ref{Mag KO beta nu eq} gives $\beta / \nu$. This ratio of critical exponents is informative, as the larger it is, the more strongly the magnetization changes as we apply $R_b$ and move away from the critical manifold. Additionally, it tells us how the magnetization and correlation length compare in their sensitivities to changes in $t$. Note that, even if we had enriched our construction of the Koopman operator (e.g. by including time delayed versions of the magnetization \cite{Takens1981}), the additional eigenvalues of $\tilde{U}_m$ wouldn't be related to other critical exponents. This is because critical exponents are determined by the flow along the relevant direction(s) of a given RG fixed point, and the 2D Ising model (as well as the 2D three-state Potts model) has a single relevant dimension, in the $h = 0$ case, for its non-trivial fixed point.


The error in estimating $\beta /\nu$ using this method, as a function of initial condition in K-space, is shown in Fig. 1a. We compared these results to those obtained by using the standard Monte Carlo RG (MCRG) method \cite{Ma1976, Swendsen1979, Swendsen1984, Newman1999a}. This method approximates the RG transformation near the fixed point Hamiltonian, $H^*$, by the linearization
\begin{equation}
\label{T linearization}
    T_{\alpha \beta} = \left[ \partial K_{\alpha}^{(n+1)} / \partial K_{\beta}^{(n)} \right]_{H^*}
\end{equation}
$\partial K_{\alpha}^{(n+1)} / \partial K_{\beta}^{(n)}$ can be solved for by using the chain rule and certain identities that require computing spin-spin correlations \cite{Niemeyer1974, Swendsen1979}. If the matrix $T$ is constructed by using only even interactions in the fixed point Hamiltonian (i.e. $K_2, K_4,$ etc.), $\nu$ is related to the largest eigenvalue of $T$. If $T$ is instead constructed by using only odd interactions (or in the case of \cite{Swendsen1979}, only $K_2$ and $h$), the critical exponent $\eta$ is related to the largest eigenvalue of $T$. The remaining critical exponents are found using standard critical exponent relationships \cite{Goldenfeld2005} (see Supplemental Material).

\begin{figure}
\centering
\includegraphics[width = 0.65\textwidth]{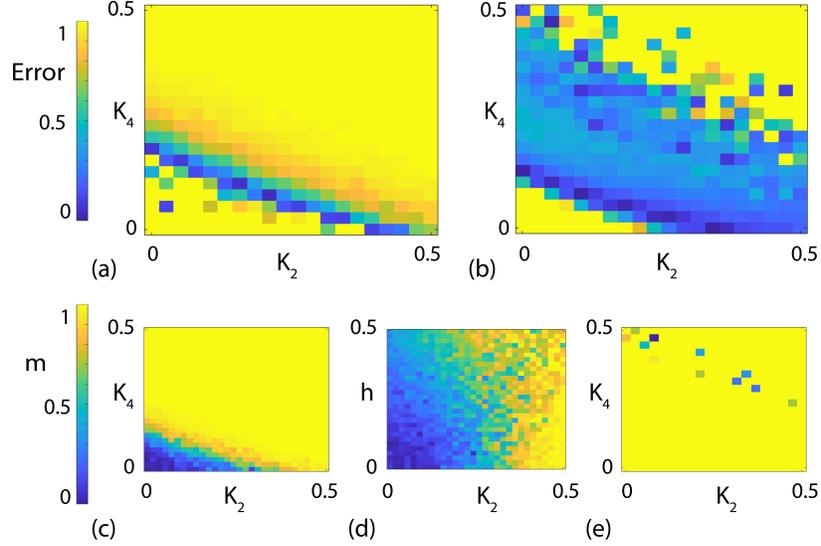}
\caption{\textbf{Error in approximating $\beta / \nu$ for the 2D Ising model.} (a) Error in approximating $\beta / \nu$ using the KOT finite section method. (b) Error in approximating $\beta / \nu$ using the MCRG method, while fixing $\eta = 0.25$ \cite{Goldenfeld2005}. This limits our computation to be over even K-space. (c) RG iteration time averaged magnetization in the $(K_2, K_4)$ plane. This serves as a proxy for the RG flow. (d) same as (c), but for the flow in the $(K_2, h)$ plane. (e) Error in approximating $\beta / \nu$ using the MCRG method, setting $h = 0$, as was done in (a). For all error subplots (a, b, and e), a maximum value of 100\% (yellow) was set. }
\end{figure}

We note that the MCRG requires knowledge of the RG flow in both the odd and even K-spaces to compute $\beta / \nu$ (as both $\eta$ and $\nu$ are needed to solve for $\beta$), whereas the finite section method does not. If we ignore some information, for instance, if we fix the odd interaction to be $0$ (as we did when we using the finite section method), there is large error over a considerable range of K-space when computing $\beta / \nu$ (Fig. 1e). This is because the flow in the odd K-space is different from that in the even K-space (Fig. 1c,d). Fixing the value of $\eta$ to it's exact value of $0.250$ \cite{Goldenfeld2005}, and searching only over even K-space, the MCRG computed $\beta / \nu$ is slightly better as compared to our Koopman RG method, but is accurate over a larger range of K-space (Fig. 1b). Similar results are seen when evaluating $\beta / \nu$ in the 2D three-state Potts model (see Supplemental Material Fig. S1). 

While we have plotted $\beta / \nu$ because of it's direct relation to the observable we recorded across RG iteration time, we can also use standard critical exponent relationships to compute another critical exponent ratio, as well as two ``full'' critical exponents (see Supplemental Material for more details). 
\begin{equation}
    \label{gamma nu eq}
    \gamma / \nu = d - 2\beta / \nu
\end{equation}
\begin{equation}
\label{eta eq}
    \eta = 2 - d + 2 \beta / \nu
\end{equation}
\begin{equation}
    \label{delta eq}
    \delta = d \left( \beta / \nu\right)^{-1} - 1
\end{equation}
where $\gamma$ is critical exponent related to the scaling of the magnetic susceptibility near $T_C$, $\eta$ is the critical exponent related to the behavior of the two point correlation function right at $T_C$, and $\delta$ is the critical exponent related to the scaling of the energy of the system as a function of the magnetization near $T_c$. Evaluating Eqs. \ref{gamma nu eq}-\ref{delta eq} using the $\beta / \gamma$ found at $K_2 = 0.3947, K_4 = 0$, and $h=0$ from 100 different simulations, gives mean values of $\beta / \nu = 0.1269$, $\gamma / \nu = 1.74632$, $\eta = 0.2537$, and $\delta = 14.7652$. These compare well to the true values $\beta / \nu = 1/8$, $\gamma / \nu = 7/4$, $\eta = 1/4$, and $\delta = 15$ \cite{Goldenfeld2005}.

Finally, our KOT method does not require any explicit calculation of how applying the block spin RG changes \textbf{K}. This property shows the power of recognizing the RG as a Koopman operator: simply by recording an observable as we apply the RG, we can recover properties of the system in the form of (ratios of) critical exponents. This removes the constraint of only being able to work on systems where solutions for the linearization of $T$, $\partial K_{\alpha}^{(n+1)} / \partial K_{\beta}^{(n)}$, are available. Therefore, the Koopman RG allows for the calculation of (ratios of) critical exponents even in systems that are not translationally invariant, as long as a coarse graining procedure exists. Because there have been some that have been successfully used (e.g. \cite{Newman1999b}), this is not a particularly strong requirement. 

\section{\label{sec:level1}Discussion \protect\\ }

We started by connecting the renormalization group to a powerful nonlinear dynamical systems theory, Koopman operator theory. This insight was suggested by the fact that both KOT and the RG have been shown to be related to the asymptotics of partial differential equations \cite{Chen1994, Chen1996, Kutz2016}, principal component analysis \cite{Brunton2016JNeuro, Klus2018, Bradde2017}, normal form theory \cite{Mauroy2016, Mohr2016, Mezic2019, Raju2019}, and machine learning \cite{Mehta2014, Kawahara2016, Takeishi2017a, Takeishi2017b}. We showed that, when viewed as a dynamical system in ``RG iteration time'', by definition, the RG is a Koopman operator in the space of coupling constants. The fact that KOT has been successfully applied to understanding the dynamics of a wide range of systems in a data-driven manner \cite{Rowley2009, Budisic2012, Mezic2013, Brunton2016, Korda2018, Hogg2019, Fonoberova2018, Georgescu2015} led us to investigate whether we could import methods from KOT to evaluate critical exponents, and whether those methods would afford us benefits that the standard Monte Carlo RG \cite{Newman1999a, Ma1976, Swendsen1979, Swendsen1984} approach cannot. 

We showed that, for both the 2D Ising model and the 2D three-state Potts model, the Koopman RG approach, which made use of the finite section method, allowed for an evaluation of $\beta / \nu$ that was close to as good as that of the MCRG on the same part of K-space, yet required less information about the exact RG flow (Fig. 1 and Fig. S1). This is because the MCRG requires knowledge of the RG flow in both the odd and the even K-spaces, whereas the KOT RG method does not. Additionally, our method was significantly faster, as we only had to measure the magnetization at each RG iteration time point, whereas the MCRG required computing spin-spin correlations of all spin pairs over a lengthy time period (see Supplemental Material). We also showed that, by using critical exponent relationships \cite{Goldenfeld2005}, we were also able to compute $\gamma / \nu$, $\eta$, and $\delta$ to good agreement with their true values.   

While the robust and fast calculation of these (ratios of) critical exponents is encouraging, the real advantage in using the Koopman RG approach is that it does not rely on any explicit formulation of the RG transformation in terms of $\mathbf{K}$, like the MCRG does. This greatly widens the range of systems that we can apply the RG framework to. Applying our method to problems that are not translationally invariant and, therefore, have been largely unstudied using numerical real space RG methods, will be a direction of future work. The example of deep neural networks is particularly exciting \cite{Mehta2014}. By computing their (ratios of) critical exponents, we can establish universality classes. Such universality classes could offer a principled way in which to explore how architectures, learning rules, and data sets affect the performance of DNNs. More generally, the perspective taken here has already inspired work on accelerating the training of DNNs via KOT \cite{Dogra2020a, Dogra2020b}. The Koopman RG could also be applied to spin glasses. Because spin glasses are not translationally invariant, MCRG based methods require a mapping to an effective, nonrandom Hamiltonian \cite{Wang1988}. This mapping is tricky, and it is not clear that it does not affect the results (e.g. Table 1 of \cite{Jorg2008}). While there is a plethora of work computing the critical exponents of spin glasses via other approaches (including many that specifically compute $\eta$), there is much disagreement (as reviewed in \cite{Jorg2008, Katzgraber2006}). Whether these differences arise from nonlinear corrections to the RG equations is something that we hope to explore using the Koopman RG. 

Finally, while our results followed from applying tools from KOT, there is significantly more to the theory than what we have used here. For example, KOT allows us to represent the coarse grained Hamiltonian as 
\begin{equation}
\label{RG decomp}
R_b H(\textbf{K}_0) = \sum_{k} \mu_k \mathbf{v}_k \phi_k \left(\textbf{K}_0\right)
\end{equation}
where $\textbf{v}_k$ is the k$^{th}$ Koopman mode, $\phi_k$ is the k$^{th}$ Koopman eigenfunction, and $\mu_k$ is the k$^{th}$ eigenvalue \cite{Mezic2019,  Williams2014}. As we did earlier, we can consider the evolution of $H(\textbf{K}_0)$ in RG iteration time. The same representation, after $n$ applications of the RG, then gives 
\begin{equation}
\label{RG decomp RG time}
R_b^n H(\textbf{K}_0) = \sum_{k} \mu_k^n \mathbf{v}_k \phi_k \left(\textbf{K}_0\right)
\end{equation}
From this, we see that the dynamics of the RG flow, in each of the directions of K-space defined by the Koopman modes, is determined by the magnitude its corresponding $\mu_k$ \cite{Williams2014}. While the accurate computation of the Koopman modes and eigenfunctions is difficult (although doable \cite{Mezic2005, Mezic2019b, Williams2014}), this representation, in principle, allows us to find the fixed points, and their relevant and irrelevant directions, of the RG flow all from measuring a single observable. Because these are not known for many systems, and because recent work has shown that even relatively simple models can have non-trivial fixed points with exotic features \cite{Young2020}, we believe this is an area the Koopman RG can play an important role in. 

Our paper highlights the power that applying techniques developed in the field of nonlinear dynamical systems offers when working with the RG. We see our work very much in the same spirit as the recent success in using normal form theory to classify nonlinear generalizations of scaling functions \cite{Raju2019}. Given the connection of KOT with normal form theory, and nonlinear systems in general \cite{Mauroy2016, Mohr2016, Mezic2019}, it would be interesting to see whether our method could be used to provide numerical predictions as to what universality families systems belong to.

KOT is an exciting, powerful, and growing theory, and we hope that this illustrates its potential as a tool for physicists.

\section*{\label{sec:level1}Acknowledgements \protect\\ } We would like to thank Prof. Igor Mezic for the introduction to, and helpful discussions on, KOT. We also thank Dean Huang for his advice on early attempts at this problem and Akshunna S. Dogra for discussions on DNNs. We thank the referees for their constructive feedback, which strengthened the paper. The author is supported by a Chancellor Fellowship from UCSB.




\end{document}